\documentclass[11pt,twoside]{article}

\usepackage{asp2006}
\usepackage{epsf}
\usepackage{lscape}

\begin{document}

\title{On the Infrared Properties of Dusty Torus}

\author{Liza Videla, \altaffilmark{1} Paulina Lira, \altaffilmark{1}
 Almudena Alonso-Herrero, \altaffilmark{2} David M. Alexander,
 \altaffilmark{3} and Martin Ward \altaffilmark{3}} 

\altaffiltext{1}{Departamento de Astronom\'ia, Universidad de Chile,
Casilla 36D, Santiago, Chile}

\altaffiltext{2}{Instituto Departamento de Astrof\'isica Molecular e
Infrarroja, Instituto de Estructura de la Materia, CSIC, E-28006,
Madrid, Espa\~na}

\altaffiltext{3}{Department of Physics, Durham University, South Road,
Durham, DH1 3LE, UK.}

\begin{abstract}
We performed imaging on 49 type 2 Seyfert galaxies in 6 near- and
mid-infrared bands (1-10$\mu$m). We are separating the contribution of
the torus from the host galaxy by radial profile fitting techniques
and we will compare the observed spectral energy distributions with
theoretical models of torus emission to constrain geometrical and
physical parameters.
\end{abstract}

\section{Introduction}
In order to explain the differences observed between type 1 and type 2
 AGN it has been proposed that an axially symmetric dusty structure
 (the torus) beyond the accretion disk absorbes a considerable
 fraction of the radiation emitted at wavelengths shorter than 1
 $\mu$m (the AGN Unified Scheme). The dust in this torus typically
 reaches temperatures of a few hundred degrees and therefore its
 emission peaks somewhere at IR wavelengths. It is therefore in this
 wavelength regime that the torus can be detected allowing us to
 determine its properties.

\section{Observations, data analysis and preliminary results}

We have obtained images for 49 Seyfert 2 galaxies in 6 IR bands: J
(1.25 $\mu$m), H (1.65 $\mu$m), K (2.2 $\mu$m), L (3.78 $\mu$m), M
(4.66 $\mu$m) and N (10.36 $\mu$m) using the VLT at Paranal, NTT at La
Silla and Gemini at Pach\'on. These galaxies were selected from the
\emph{Extended 12 $\mu$m Galaxy Sample} \citep{rush}.

We obtained the surface brightness profiles of each galaxy and modeled
them to separate the contribution from a nuclear point source and a
stellar component (disk+bulge, Figure 1) to construct the nuclear
spectral energy distributions (SEDs) for each galaxy. The SEDs are
then fitted using theoretical models of the emission of the dusty
torus from \citet*{nenkova}, who solved the radiative transfer problem
of a clumpy torus including absorption, emission and scattering.

In Figure 2 we show results for MCG -2-40-004 and MCG -3-34-64, both
galaxies classified as type 2 Seyfert galaxies in \cite{rush}, but in
\cite{veron} MCG -2-40-004 is classified as a type 1.9 Seyfert, and
MCG -3-34-64 as a type 1 due to the detection of broad polarized
Balmer lines. Following \citet{nenkova} definitions, for MCG -2-40-004
we obtained a torus of 100 pc in size, with 5 clouds along the line of
sight (each with a $\tau_\nu$=20) and an inclination of 60$\deg$. For
MCG -3-34-64 we obtained a torus of 30 pc in size, with 5 clouds along
the line of sight (each with a $\tau_\nu$=40) and a face on
inclination. This SED is clearly more consistent with a type 1 Seyfert
classification.

\begin{figure} 
\plottwo{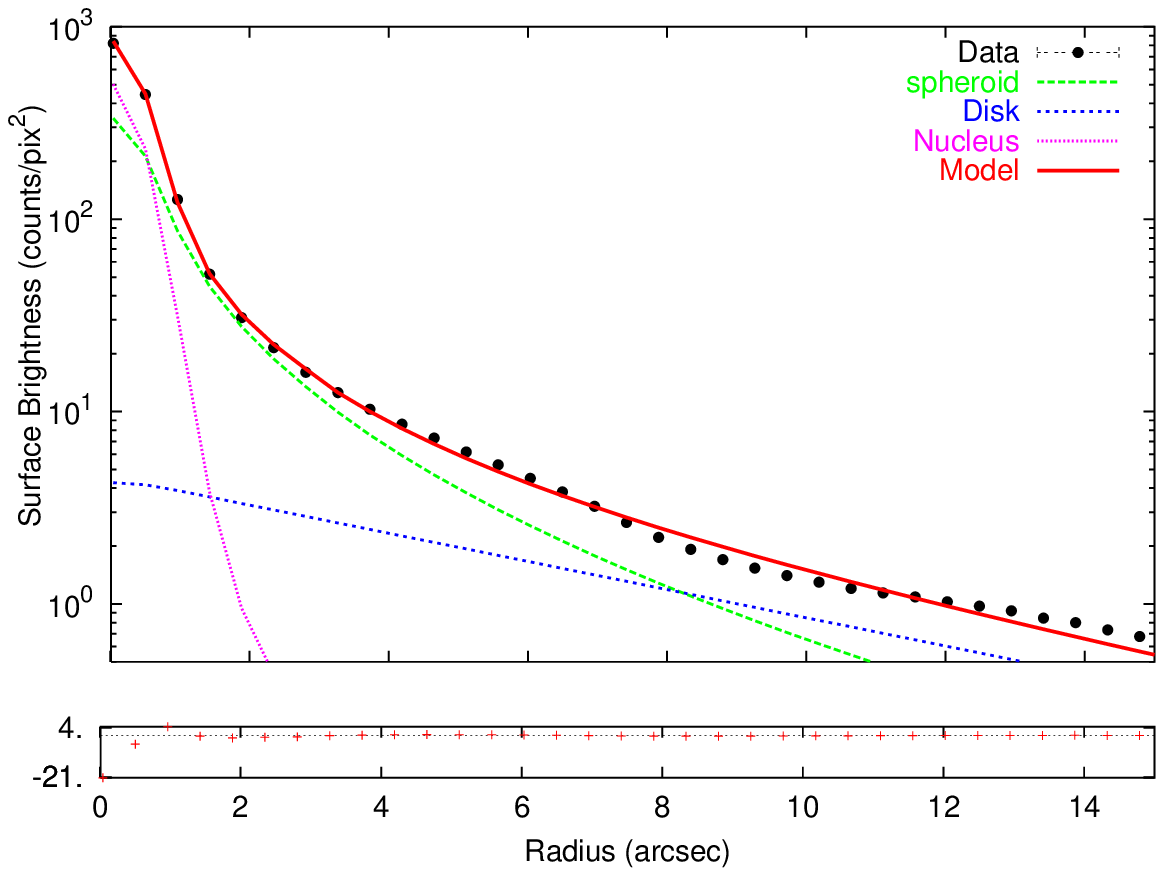}{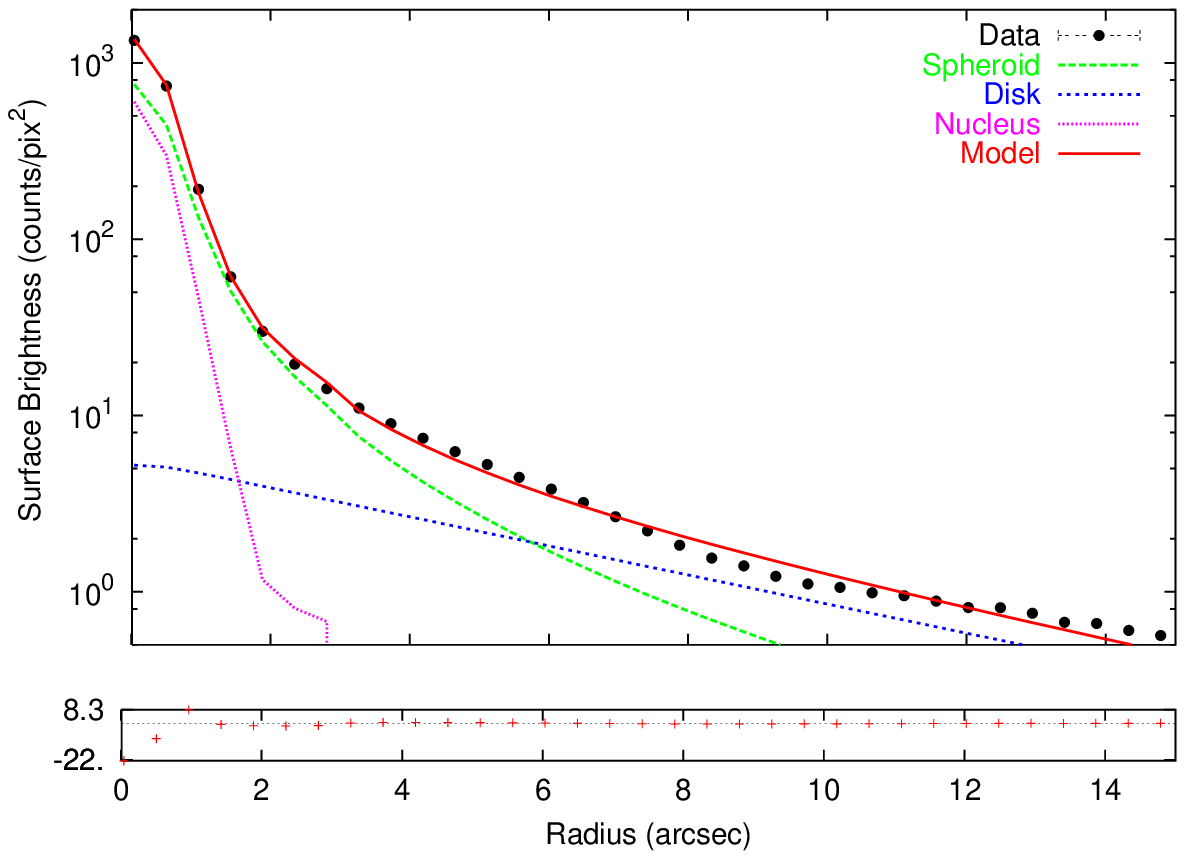}
\label{fig1}
\caption{Surface brightness profile of MCG -2-40-004 at H (left) and K
   (right) bands. Dots represent the data points, the dashed line the
   disk component, the dot-dashed line the bulge component, and the
   doted line the nuclear contribution. The solid line is the sum of
   the 3 components.}
\end{figure}

\begin{figure} 
\plottwo{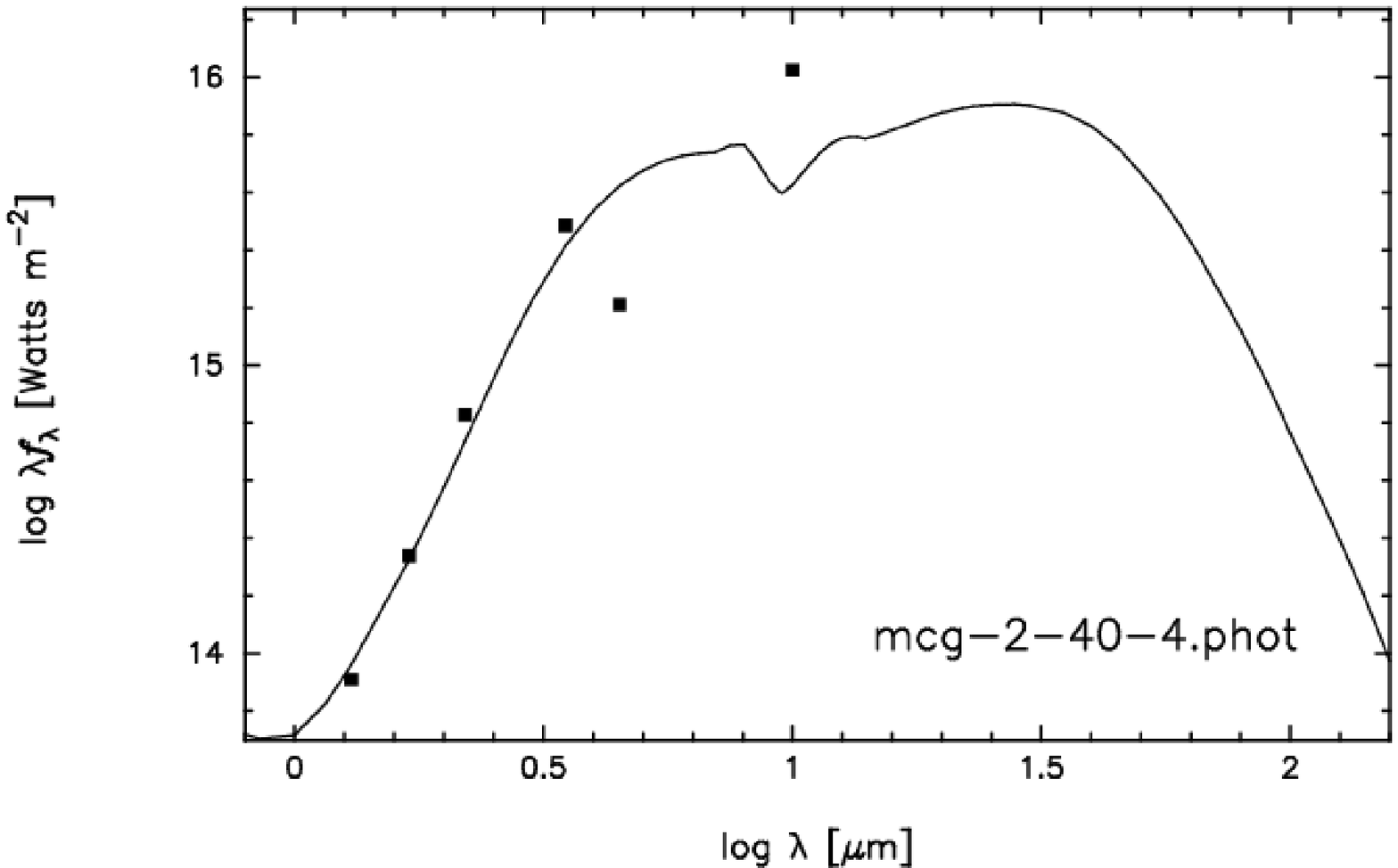}{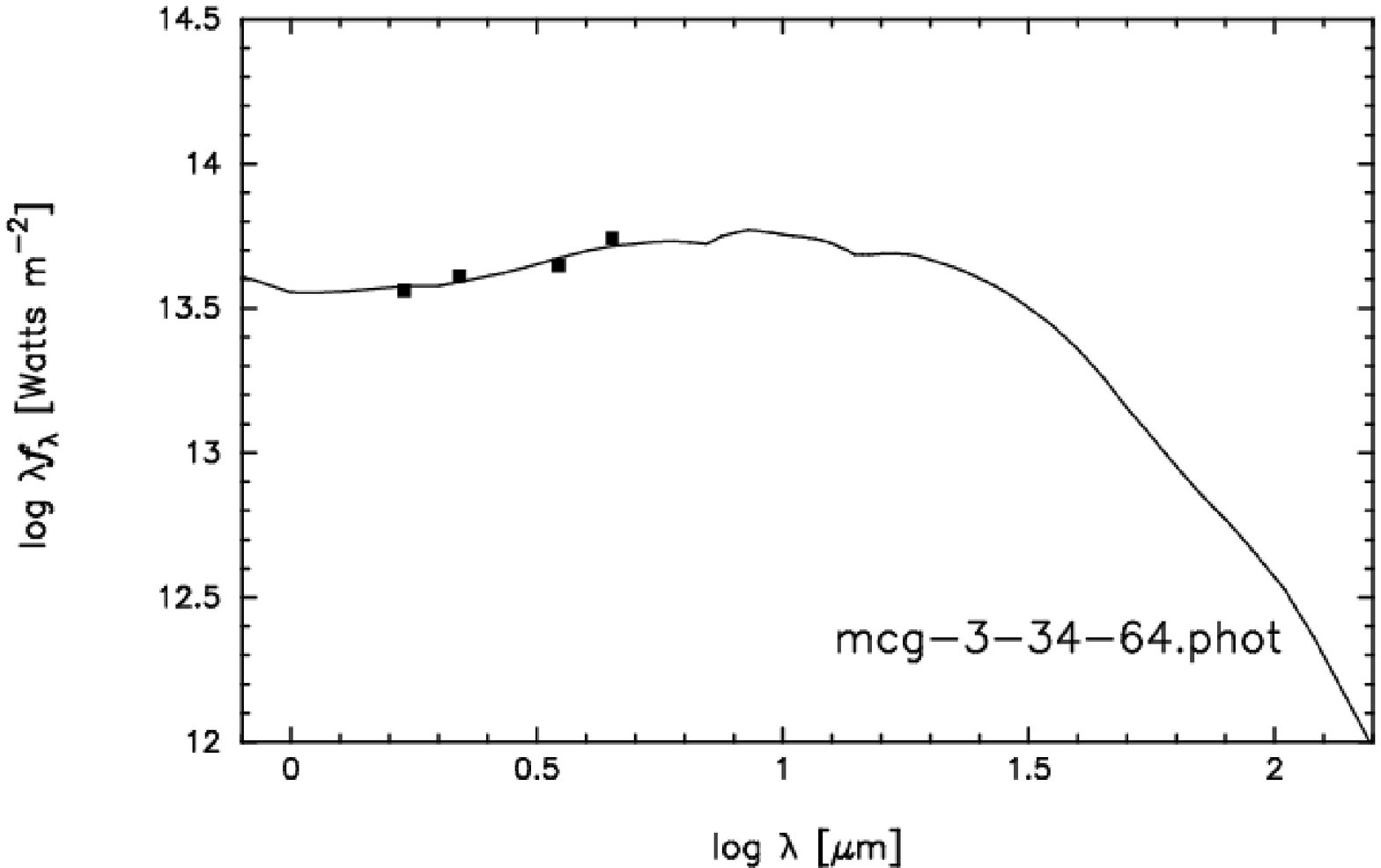}
\label{fig2}
\caption{Nuclear SEDs of MCG -2-40-004 (left) and MCG -3-34-64
(right), and their fitted model.}
\end{figure}

Complementary data for our study comes from X-ray, spectropolarimetry,
ISOCAM and Spitzer observations. Together with our observations, we
will use this information to further constrain the SEDs of our targets
and determine the geometrical and physical parameters that govern the
IR emission from AGNs.

\acknowledgements{LV thanks support by Conicyt to attend this
meeting.}

\end{document}